\documentstyle{aipproc}

\begin{document}
\title{Gamma-Ray Bursts from Extragalactic Magnetar Flares}

\author{Robert C. Duncan}
\address{Dept.~of Astronomy, University of Texas, Austin, Texas 78712}

\def\lessim{\lower 0.5ex\hbox{${}\buildrel<\over\sim{}$}}
\def\gtrsim{\lower 0.5ex\hbox{${}\buildrel>\over\sim{}$}}
\def\VVmax{\langle V/V_{max} \rangle}

\maketitle

\begin{abstract}

The prototype for events that we call MFs---``March Fifth" events 
or ``Magnetar Flares"---was observed on March 5, 1979.  \ 
There is evidence that MFs are powered by catastrophic 
magnetic instabilities in ultra-magnetized neutron stars.  These events 
begin with brief ($\Delta t\sim 0.1$--1 s), intense, hard spikes of 
gamma rays, probably emitted in concurrence with relativistic outflows; 
followed by long ($t \sim 100$ s) softer tails of hard X-rays, modulated
on the stellar rotation period.  Prototypical MFs could have been 
detected by BATSE out to $\sim 13$ Mpc, nearly reaching the Virgo cluster.  
The likely number of isotropic, standard-candle MFs detected 
by the BATSE experiment is $\sim 12$.  
These short-duration, fast-rising gamma-ray
bursts could in principle be identified by their positional coincidences with 
nearby galaxies and the Supergalactic Plane.  The ensuing soft tail 
emission would not have been detected by BATSE for sources more distant than 
the Andromeda Galaxy.  Bayes' Theorem
implies that there is a $\sim 99\%$ chance for at least 1 isotropic
MF in the BATSE catalog, and a $\sim 16\%$ chance for more than 20. \ 

 It is possible that MFs also emit an intense, hard, {\it beamed} component 
during the intial spike phase.  
If this beamed component has opening angle $\psi = 8^\circ \psi_8$ and 
peak luminosity 
comparable to the power of the isotropic component, then BATSE would 
detect such  beamed sources out to redshift $z \sim 0.1 \, \psi_8^{-1}$, 
at a full-sky rate of \hbox{$\dot{N}_B \sim 100 \, \psi_8^{-1}$ yr$^{-1}$.}  
We speculate that such beamed MFs could account for the short, hard Class II 
gamma-ray bursts (GRBs) in the BATSE catalog, or some significant 
subset of them.  
If true, then Class II GRBs positions should correlate with the positions
of  galaxies and galaxy clusters within $\sim 350$ Mpc.  

\end{abstract}

\section*{Introduction}

    The short-duration, hard-spectrum gamma-ray bursts (GRBs) in the BATSE 
catalog, often called Class II bursts, have no detected  afterglows. 
As a consequence, they have not yet been subject to scrutiny for absorption 
lines, and the distance from Earth at which these 
bursts originate is uncertain.  Here we will explore the 
possibility of a relatively nearby extragalactic origin, $z \lessim 0.1$,
for (at least) some Class II GRBs. 

Kouveliotou et al.~first showed that the GRB population is
bimodal \cite{kou93}.  More recently, Murkherjee et al.
found evidence that there exist {\it three} classes of 
BATSE GRBs \cite{mur98}.  But Hakkilla et al.~(ref.~\cite{hak99}) argued 
that Class III properties can be produced from Class I by a 
combination of measurement error, hardness-intensity correlation, and a 
newly-identified BATSE bias, the fluence duration bias.  Class II GRBs,
which are short-duration, low-fluence, hard-spectrum events, do not seem to 
be related to Class I/III. \ It is Class II GRBs, or some 
subset of Class II, that we suggest are MFs.  

  Class II GRBs have many features in common with the hard spikes of
MFs. \ About $40\%$ of Class II bursts are single-peaked \cite{bha99}. 
Most peaks exhibit a fast rise and
slower decline, with hard-to-soft spectral evolution \cite{bha99}.  These
properties are shared by the hard spikes of the March 5th and August 27th 
events \cite{maz79,maz99,fer00}, which also had similar spectral hardness
\cite{fen96,fkl96}.  
Class II GRBs that are sufficiently bright to study with fine time
resolution often show substructure down to time scales of
$\sim 10$ ms and less \cite{mee94,mee96}, as did the 1979 March 5
event \cite{bar83}.

\section*{BATSE detection of MF's}

Consider a beamed source of gamma-rays, emitting uniformly into a 
beaming fraction $f = \Delta \Omega / 4 \pi,$
of the unit sphere, with peak beam luminosity  $L_{\rm bm}$. 
For a BATSE peak-flux detection threshold $F_B$, the sampling
depth of BATSE is 
\begin{equation}
D =\left({ L_{\rm bm} \over 4 \pi f F_B}\right)^{1/2} .
\label{batdepth}
\end{equation}
If $D$ is small enough that source density evolution and departures
from Euclidean geometry are negligible, then
BATSE's rate of event detection is 
\begin{equation}
\dot{N}_B = {4 \pi\over3} \, D^3 n_* f \, \Gamma .  
\end{equation}
Here $n_*$ is the density of $L_*$ galaxies (with 
luminosities comparable to that of the Milky Way) and $\Gamma$
is the rate of MFs within each $L_*$ galaxy.  
Thus the (full-sky) BATSE detection rate is
\begin{equation} 
\dot{N}_B = {n_* \, \Gamma \over 6 \, (\pi f)^{1/2}} \ 
\left({ L_{\rm bm} \over F_B }\right)^{3/2}.
\label{batrate}
\end{equation}

 We evaluate this using known parameter values.
In particular, $F_B \approx 10^{-7}$ erg cm$^{-2}$ s$^{-1}$  
(ref.~\cite{mee94}) 
and $n_* =  0.01 \, h^3$ Mpc$^{-3}$  from the local normalization of the 
Schechter luminosity function (e.g., ref.~\cite{pee93}).
We will adopt a Hubble constant of $H_o = 65 \, h_{65}$ km s$^{-1}$ 
Mpc$^{-1}$, thus $n_* = 2.8 \times 10^{-3} \, h_{65}^3 \ \hbox{Mpc}^{-3}. $

Given 2 MFs in our Galaxy (or in nearby dwarf satellite galaxies)
in the past $t_o \sim 20$ years of effective full sky coverage 
during which have had capability for 
detecting them (Hurley 1999, private communication), the MF rate per 
$L_*$ galaxy is roughly
\begin{equation}
\Gamma = 0.1  \ \Gamma_{0.1}  \ \hbox{yr}^{-1}.
\label{gamma}
\end{equation}

The simplest idealization is that the gamma-rays in MF hard spikes are 
emitted isotropically, $f \approx 1$.  Since the long, soft, oscillating 
tails of MFs have
never been observed without hard spikes at their onsets, we infer
that many or most MFs begin with a spike of quasi-isotropic emissions. 
(This is in addition to a possible high-intensity, beamed component. 
We have no direct evidence for 
beamed emission; however, with only 2 detected MFs, we would not 
expect to have observed any beams with $f\ll 1$.)
 
Setting $f = 1$ in the above formulae, we find a BATSE sampling depth
for the isotropic component of MFs of 
\begin{equation}
D= 13 \, (L_{45}/2)^{1/2} \ \hbox{Mpc}, \, 
\label{diso}
\end{equation}
where we have scaled to the value of peak luminosity found for the
1979 March 5 event by Fenimore \cite{fen96}.  This falls just short of 
the Virgo cluster at  $D[{\rm \scriptstyle Virgo}] \approx 18 \, 
h_{65}^{-1}$ Mpc.  The  full-sky BATSE detection rate is 
\begin{equation}
\dot{N}_B = 2.6  \, (L_{45}/2)^{3/2} \, \Gamma_{10} \ h_{65}^3 \
\hbox{yr}^{-1} 
\end{equation}
Since BATSE operated for 9.5 years, corresponding to 4.75 yrs of full-sky
coverage, there should be 
\begin{equation}
\langle {\cal N}\rangle \sim 12 \ (L_{45}/2)^{3/2} \ 
\Gamma_{0.1} \ h_{65}^3 
\label{niso}
\end{equation}
extragalactic MFs in the BATSE catalog, detected via their 
quasi-isotropic hard spike emissions.  

The most uncertain parameter in this estimate is 
$\Gamma$, the galactic rate of MFs.  
The Bayesian probability distribution for $\Gamma$, 
${\cal P} \equiv dP/d\Gamma$, is 
\begin{equation}
{\cal P} (\Gamma)  = { P( 2\, | \, \Gamma) \ {\cal P}_{prior}(\Gamma) \over
\int_0^\infty d\Gamma \ P( 2 \, | \, \Gamma) \ {\cal P}_{prior}(\Gamma)},
\end{equation}  
where the probability for observing 2 MFs  in time $t_o\approx 20$ yr, 
given $\Gamma$, is 
$P(2 \,| \, \Gamma) = {1\over2} (\Gamma t_o)^2 \,  \hbox{exp}(-\Gamma t_o)$ 
since this is a Poisson process. \ The appropriate prior distribution, 
when we don't know the order of magnitude of $\Gamma$ {\it a priori}, is 
$(d P_{prior}/d \hbox{log}\Gamma) = \hbox{constant}$, or 
${\cal P}_{prior} \propto \Gamma^{-1}$.  \  Thus
\begin{equation}
{\cal P}(\Gamma) = \Gamma \,  t_o^2 \ \hbox{exp}(-\Gamma t_o),
\label{probdist}
\end{equation}
 with a mean value $\langle \Gamma \rangle = 2 \, t_o^{-1} = 0.1$ yr$^{-1}$ 
as noted above [eq.~(\ref{gamma})].  
The probability for the galactic flare rate to exceed a cutoff value, 
$\Gamma > \Gamma_x$, is thus 
\hbox{$P(\Gamma > \Gamma_x) = \hbox{exp}(-\Gamma_x \, t_o) \, 
(1 + \Gamma_x \, t_o)$}.
If other sources of uncertainty can be neglected, then
there is a $99\%$ probability
for at least one isotropic MF in the BATSE catalog, ${\cal N} > 1$.
\ The probability is $80\%$ for ${\cal N}>5$; 
$52\%$ for ${\cal N}>10$; $16\%$ for ${\cal N}>20$; 
and $4.5\%$ for ${\cal N}>30$.

\section*{Could all Class II GRB's be MF's?}

Suppose that there also exists a beamed component in the intial hard spikes of
MFs, with a full opening angle $\psi$.
The beaming fraction is $f = [1 - \cos (\psi/2)]$ assuming two beams, one at 
each magnetic pole.  This is
$f \approx \psi^2/8$ for $\psi \ll1$, or 
\hbox{$f = 2 \times 10^{-3} \, \psi_8^2$,}
where $\psi_8\equiv (\psi / 8^\circ)$.  From 
eqs.~(\ref{batdepth})--(\ref{batrate}), the BATSE sampling depth for beamed
MFs is
\begin{equation} 
D = 350 \ \psi_8^{-1} \ 
\left({L_{\rm bm}\over 3 \times 10^{45} \, \hbox{erg s}^{-1}}\right)^{1/2}
\ \hbox{Mpc},
\label{beamdepth}
\end{equation}
and the (full-sky) rate of detection is
\begin{equation}
\dot{N}_B = 100 \  \Gamma_{0.1} \ h_{65}^3 \ \psi_8^{-1} \
\left({L_{\rm bm}\over 3 \times 10^{45} \, \hbox{erg s}^{-1}}\right)^{3/2}
\hbox{yr}^{-1} 
\label{beamrate}
\end{equation}

This is plausibly 
in agreement with the rate of detection of Class II bursts by BATSE.  
Out of 796 bursts
classified by Murkherjee et al.,  about 185 were Class II, or $\sim 23\%$.
Since BATSE detects about 300 GRB per year with half-sky coverage
(due to the fact that the {\it Compton Observatory} is in low Earth orbit), 
the full-sky detection rate of Class II bursts is roughly \ 
$\dot{N}_B (\hbox{Class II})\approx 140 \ \hbox{yr}^{-1}. $  

However, there is no compelling reason to expect such a beamed emission 
component based upon the magnetar model\cite{dt92,td95,td01,fer00}.  

Moreover, the number-intensity or $V/V_{\rm max}$ distribution of Class II GRBs 
seems to give evidence against a local extragalactic origin for these events. 
This is a serious concern; however, note that the evidence for $\VVmax < 0.5$ in 
Class II bursts is much less compelling than in Class I/III.  It is possible
that selection effects cause a paucity of BATSE events just above
threshold for peak flux on 256 and 64-ms time scales.
(Flux averaged over 1.024 s does not give a good brightness measure for
these short bursts.)  Faint and poorly-measured bursts 
are systematically removed because of insufficient 
information to make class identifications (e.g., only 778 out of 1122
catalogued bursts were classified in ref.~\cite{hak99});  and the 
class identifications themselves are least reliable near threshold 
where measurements are most statistically dubious.  
It is also possible that the {\it bright} Class II bursts include 
a contamination of physically-distinct, non-Euclidean (low $V/V_{max}$) events; 
e.g., a tail of Class I bursts extending to 
short durations, or a subclass of short bursts from galactic SGRs or AXPs 
like the bright, hard-spectrum events with $T_{90} \sim 1$ s already 
identified from SGR 1900+14 \cite{woo99}. Note that Tavani \cite{tav98} found 
\hbox{$\langle V/V_{max} \rangle = 0.458 \pm 0.044$} for short duration, soft
spectrum bursts in the BATSE catalog (66 events with $T_{90} < 2.5$ s and
$H^e_{32} < 3$), and Cline, Matthey \& Otwinowski \cite{cli99} found
\hbox{$\VVmax = 0.52 \pm 0.06$} for all BATSE bursts with $T_{90} < 0.1$ s.

\section*{Conclusions: Observational Tests}

  Under the idealization that MF hard spikes are emitted quasi-isotropically,
we find that the BATSE catalog contains $\sim 12$ 
extragalactic MFs [eq.~(\ref{niso})]. These events are expected to be
fast-rising ($t_{rise} \lessim 1$ ms), short-duration bursts,
correlated in position with galaxies at $D_{gal}< 20$ Mpc.  Insofar
as the quasi-isotropic component of MF hard spikes have uniform peak 
luminosities, 
there will be a diminishment of the peak flux with $D_{gal}$ according to
$F_{peak} \simeq L_{\rm iso} / (4 \pi D_{gal}^2)$, affording a possible
auxiliary check on associations.  As a group, these ``isotropic MFs" should 
tend to concentrate toward the supergalactic plane \cite{deV53}.  
Hartmann, Briggs \& Mannheim (ref.~\cite{har96}) found no significant 
supergalactic anisotropy in the BATSE catalog, but their statistics 
were dominated by Class I bursts.  At distances less than 
$D_{\rm vir} \sim 18 h_{65}^{-1} \ \hbox{Mpc}$
the distribution of candidate MFs may 
show a significant Virgocentric dipole moment (e.g., Fig.~3.3 
in ref.~\cite{pee93});
and a correlation with Virgo's discrete position on the sky if the 
sampling depth (eq.~[\ref{diso}]) extends as far as $D_{\rm vir}$. 

  This prediction of ``isotropic MFs" in the BATSE catalog is 
based upon direct, reliable extrapolation from observations of
the 1979 March 5 and 1998 August 27 events.  
More speculatively, in any model of MFs, a full-sky BATSE rate $\dot{N}_B$
is possible if there exists a {\it beamed emission component} with full opening 
angle 
\begin{equation}
\psi = 8^\circ \ \Gamma_{0.1} \ h_{65}^3 \ 
\left({L_{\rm bm}\over 3 \times 10^{45} \, \hbox{erg s}^{-1}}\right)^{3/2} 
\ \left({\dot{N}_B \over 100 \, \hbox{yr}^{-1}}\right)^{-1} , 
\label{opeangle}
\end{equation}
where $L_{\rm bm}$ is the total beam peak power.  This equation assumes two
polar beams; for a single beam, $\psi$ goes up by $\sqrt{2}$. \ 

If such beamed MFs accounted for many or all Class II GRBs, then the 
events would have positions 
correlated with galaxies and galaxy clusters within the BATSE sampling depth 
at redshift 
\begin{equation} 
z_{\rm \scriptscriptstyle B} = 0.076 \ \psi_8^{-1} \  h_{65} \ 
\left({L_{\rm bm}\over 3 \times 10^{45} \, \hbox{erg s}^{-1}}\right)^{1/2}.
\label{beamredshift}
\end{equation}
Several studies have found correlations of GRB positions with 
galaxy clusters in the Abell, Corwin \& Olowin (hereafter ACO) 
catalog \cite{abe89}, but only
at a statistically marginal level \cite{mar97,kol96,hur97}.
These studies have
focused primarily on bursts with small positional error boxes, which are
mostly bright Class I events. Note that the MFs 
come from young neutron stars in star-forming regions. 
Such sources are not expected to concentrate strongly within rich galaxy 
clusters,  which contain mostly early-type (gas-stripped) galaxies. 

The positional correlations expected for MFs are difficult to study using BATSE data. 
Class II GRBs, with short durations and relatively low fluences,
tend to have poorly-determined positions, often with BATSE error boxes
of size $\sim 10^\circ$.   Future experiments will localize short-duration GRBs 
well enough to test for correlations with nearby galaxies and galaxy clusters, 
hopefully making the identification of extragalactic MFs possible. 

\acknowledgments

We thank R. Knill-Dgani for discussions.
This work was supported by NASA grant NAG5-8381; by 
the Texas Advanced Research Program grant ARP-028.

\end{document}